\documentclass{article}
\usepackage{spconf,amssymb,amsmath,graphicx}
\usepackage{subcaption}
\usepackage{hyperref}

\title{Detecting Invasive Ductal Carcinoma With Semi-Supervised Conditional GANs}
\name{Jeremiah W. Johnson\thanks{The author gratefully acknowledges NVIDIA Corp for GPU donation to support this research.}}
\address{Applied Engineering \& Sciences\\
         University of New Hampshire\\
         Manchester, NH 03101}

\begin{document}
%
\maketitle
\begin{abstract}
Invasive ductal carcinoma (IDC) comprises nearly 80\% of all breast cancers. The detection of IDC is a necessary preprocessing step in determining the aggressiveness of the cancer, determining treatment protocols, and predicting patient outcomes, and is usually performed manually by an expert pathologist. Here, we describe a novel algorithm for automatically detecting IDC using semi--supervised conditional generative adversarial networks (cGANs).The framework is simple and effective at improving scores on a range of metrics over a baseline CNN.
\end{abstract}
\section{Introduction}\label{sec:intro}

Invasive ductal carcinoma (IDC) comprises nearly 80\% of all breast cancers, making it the most common phenotypic subtype \cite{desantis2011breast}. The aggressiveness of a sample is usually determined by performing a visual analysis of tissues slides from regions where the carcinoma has been detected. As such, the detection of invasive ductal carcinoma is a necessary preprocessing step for determining aggressiveness, treatment protocols, and predicting patient outcomes. Done manually, this is a time-consuming and challenging process, as it involves the pathologist scanning large regions of mostly healthy tissue to identify and delineate the relatively smaller regions of IDC. Because precise delineation of the IDC is a critical factor in the assessment of the aggressiveness of the malignancy, there is a significant need for highly accurate automatic methods for detecting IDCs.

\begin{figure}
    \begin{center}
        \includegraphics[width=0.4\textwidth]{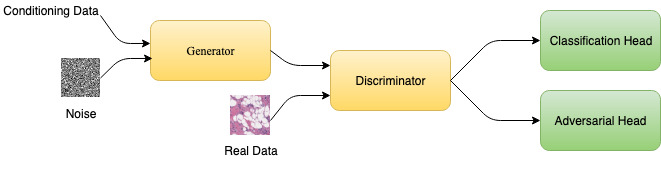}
    \end{center}
    \caption{cGAN architecture for semi--supervised training. In this case, the conditioning data is the class of the data to be generated by the generator; either IDC or healthy. Note that in contrast to the classical cGAN framework, the discriminator is not provided access to the conditioning data.}\label{fig:semi-supervised}
\end{figure}

There exist many algorithms that have been somewhat successful at automatic detection of IDCs \cite{cruzroa2014automatic, araujo2017classification}. Over the past few years, methods from deep learning, especially convolutional neural networks (CNNs), have been at the forefront of investigations into automatic detection of IDC in histopathology images \cite{cruzroa2014automatic,janowczyk2016deep}. A convolutional neural network, in general, consists of a sequence of linear and nonlinear transformations that transforms the input data into a set of features (a `learned representation') suitable for the task at hand \cite{goodfellow2016deep}. Convolutional neural networks were designed for classifying images, and the performance of CNNs on a range of challenging tasks in computer vision is state-of-the-art, often meeting or exceeding human performance \cite{krizhevsky2012imagenet,simonyan2014vgg,he2015deep,he2015delving,he2017maskrcnn}. Moreover, CNNs require little in the way of manual feature engineering, typically the most time--consuming and difficult aspect of machine learning: aside from minor preprocessing steps, the model learns the features necessary for the task at hand via the training process, which is typically a variant of stochastic gradient descent.  

Generative Adversarial Networks (GANs) were introduced in 2014 \cite{goodfellow2014gan}. A GAN consists of a pair of models, a generator and a discriminator, who compete in a minimax game: the generator attempts to generate synthetic data that is sufficiently similar to real data to fool the discriminator, and the discriminator tries to distinguish real data from synthetic data. By trading off the training process, the networks each improve until a Nash equilibrium is reached. 

GANs are often thought of as generative models, but they can be used in other ways, including for classification tasks. For example, the discriminator in a GAN can be augmented with a second network head in order to predict not only whether input data is real or generated, but also to predict the class into which the input data falls. In this regime, the generator serves to augment the existing dataset by providing the discriminator with additional synthetic training data \cite{salimans2016improved}.

In this paper, we describe a novel algorithm for automatic detection of IDC in histopathology images. The proposed algorithm uses a GAN framework where the discriminator is trained to identify IDC in both real and synthetic generated data and to distinguish real from synthetic data. The generator in the GAN framework is conditioned by class. The framework is simple and effective at improving scores on a range of metrics over a baseline CNN. The outline of this paper is as follows: Section \ref{section:background}, provides technical background on the model, while in Section \ref{section:method} the data, the methodology, and the experiments carried out are detailed; in Section \ref{section:conclusions} we present conclusions and paths for future work.

\section{Background}\label{section:background}

\begin{figure}[t!]
    \centering
    \begin{subfigure}[t]{0.15\textwidth}
        \centering
        \includegraphics[width=0.98\textwidth]{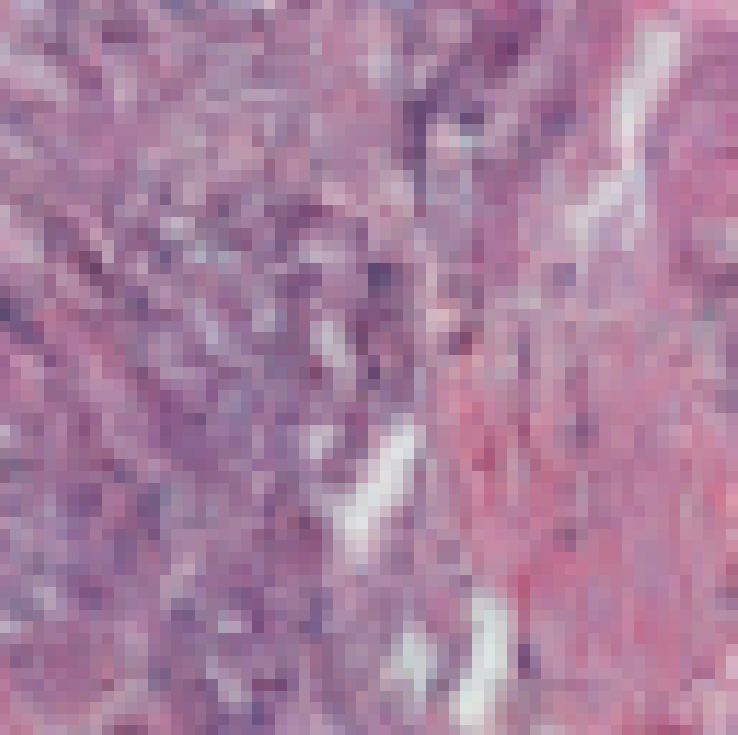}
    \end{subfigure}%
    \begin{subfigure}[t]{0.15\textwidth}
        \centering
        \includegraphics[width=0.98\textwidth]{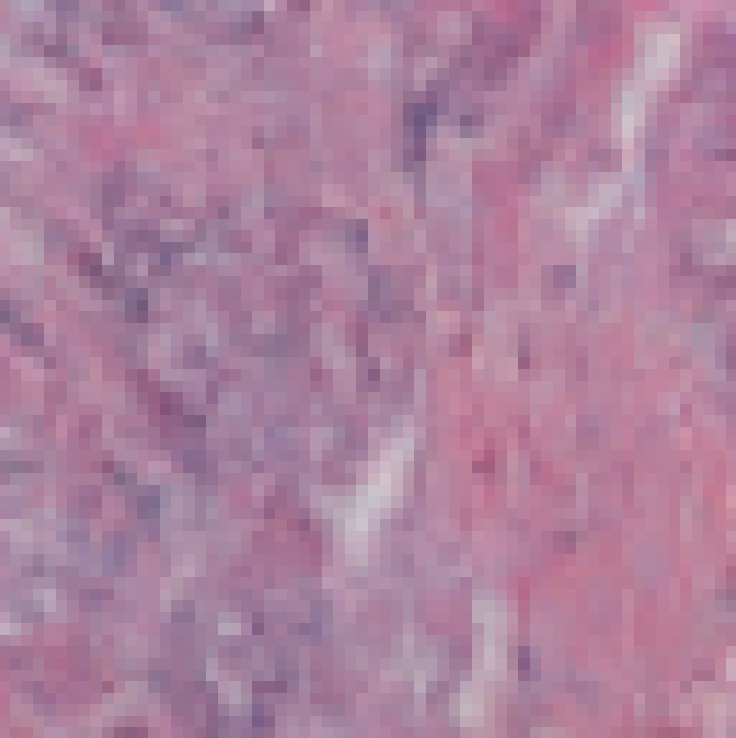}
    \end{subfigure}%
    \begin{subfigure}[t]{0.15\textwidth}
        \centering
        \includegraphics[width=0.98\textwidth]{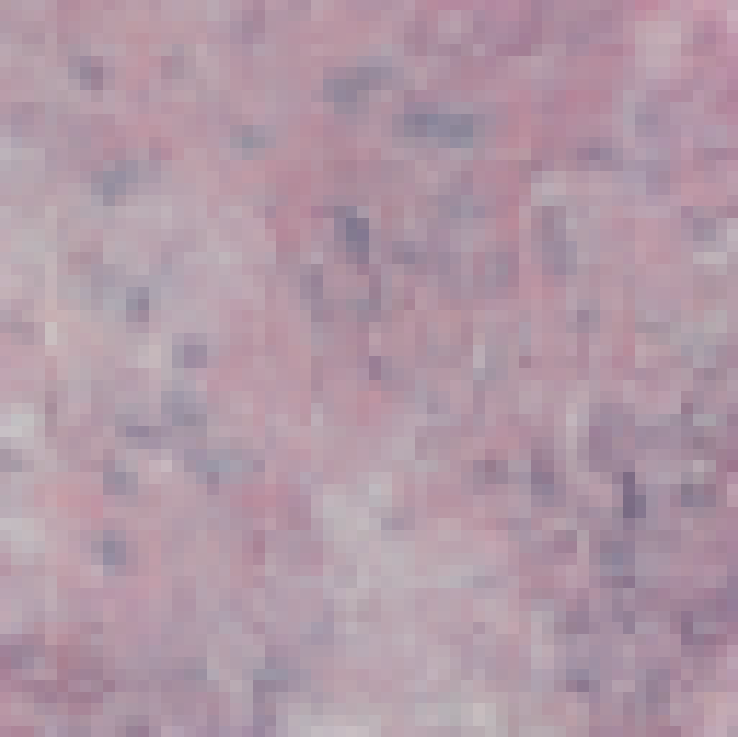}
    \end{subfigure}%

    \begin{subfigure}[t]{0.15\textwidth}
        \centering
        \includegraphics[width=0.98\textwidth]{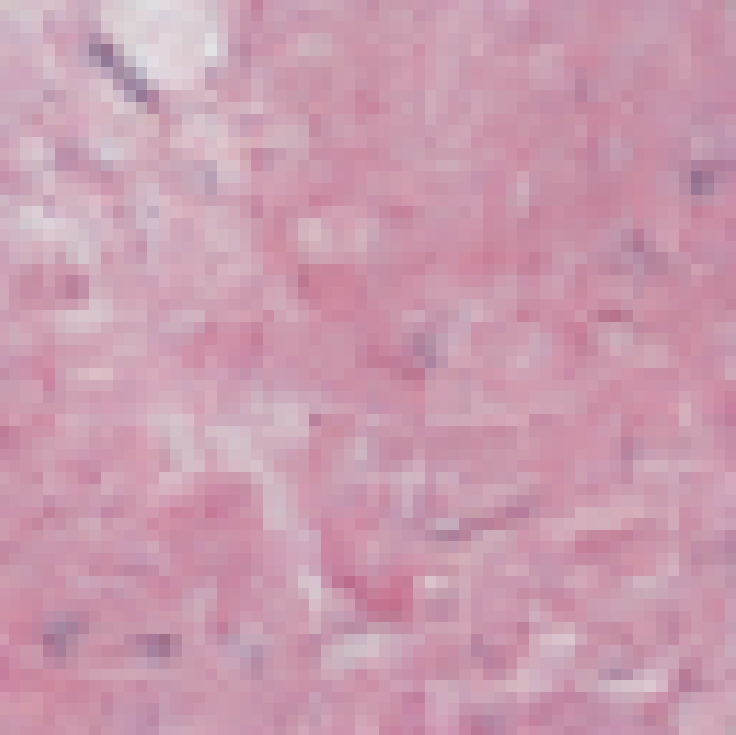}
    \end{subfigure}%
    \begin{subfigure}[t]{0.15\textwidth}
        \centering
        \includegraphics[width=0.98\textwidth]{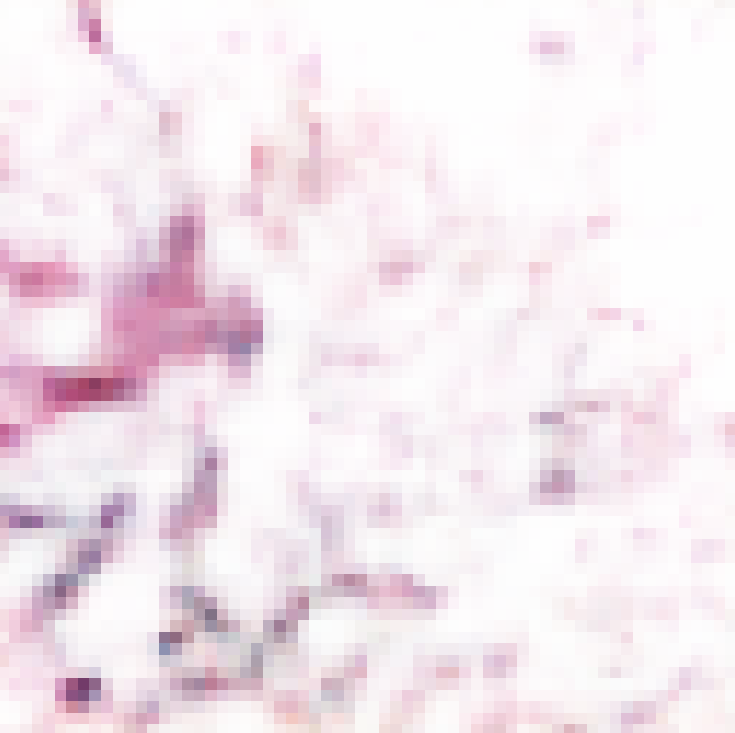}
    \end{subfigure}%
    \begin{subfigure}[t]{0.15\textwidth}
        \centering
        \includegraphics[width=0.98\textwidth]{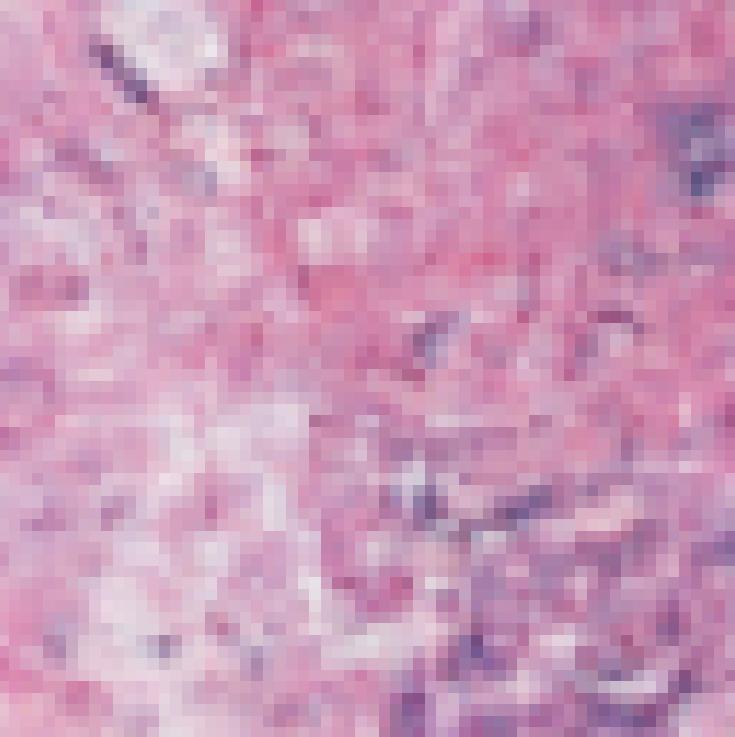}
    \end{subfigure}%

    \caption{Examples of images generated by the conditional GAN during the training process. The images in the top row were generated by conditioning the generator on IDC, while those in the second row are images produced by the generator when conditioned on healthy tissue.}\label{fig:generated-samples}
\end{figure}

\subsection{Generative Adversarial Networks}

A generative adversarial network, or GAN, consists of two neural network models, a \emph{generator} $G$ and a \emph{discriminator} $D$, that compete in an adversarial game: the task of the generator is, given some random input $\mathbf{z}$, to produce an output $G(z)$ such that the discriminator $D$ cannot distinguish $G(z)$ from a sample taken from the source domain. As $D$ and $G$ are trained in turn, $G$ learns to model the true distribution $p$ of the source domain and $D$ learns to evaluate the divergence between $p$ and the generative distribution $q$, resulting in a competition to reach a Nash equilibrium that can be expressed by the training procedure. The value function for this minimax game is given in Equation \ref{eq:gan-loss} below. 
\begin{equation}\label{eq:gan-loss}
    \begin{split}
        \underset{D}{\min}\text{\hspace{0.5em}}&\underset{G}{\max}\text{\hspace{0.5em}}V(G,D) := \\
                                               &\mathbb{E}_{\mathbf{x}\sim p}[\log(D(\mathbf{x})] + \mathbb{E}_{\mathbf{x}\sim q}[\log(1 - D(\mathbf{x}))],
    \end{split}
\end{equation}

\noindent or, equivalently,
\begin{equation}\label{eq:gan-loss-z}
    \begin{split}
        \underset{D}{\min}\text{\hspace{0.5em}}&\underset{G}{\max}\text{\hspace{0.5em}}V(G,D) := \\ 
                                               &\mathbb{E}_{\mathbf{x}\sim p}[\log(D(\mathbf{x})] + \mathbb{E}_{\mathbf{z}\sim p_{\mathbf{z}}}[\log(1 - D(G(\mathbf{z})))],
    \end{split}
\end{equation}

\noindent where $\mathbf{z}\sim p_{\mathbf{z}}$ is noise. 

GAN training is known to often be unstable and prone to issues such as mode collapse, but in recent years several notable developments including spectral normalization and gradient penalty have significantly improved the stability of GAN training \cite{miyato2018spectral,gulrajani2017improved}.

\subsection{Conditional GANs}

A conditional GAN, or cGAN, is a GAN designed to incorporate conditional information \cite{mirza2014conditional}. cGANs have been shown to be effective tasks such as clas--conditional image synthesis and image--to--image translation; in these cases, both the generator and the discriminator are provided the conditional information, usually via concatenation with the input data, though other methods have been proposed and shown to be more effective in specfic contexts \cite{isola2017image,mirza2014conditional,miyato2018cgans,mahmood2019structured}. The value function for a cGAN is given below, where $\mathbf{y}$ represents the conditioning data.

\begin{equation}\label{eq:cgan-loss}
    \begin{split}
        \underset{D}{\min}\text{\hspace{0.5em}}&\underset{G}{\max}\text{\hspace{0.5em}}V(G,D) := \\
                                               &\mathbb{E}_{\mathbf{x}\sim p}[\log(D(\mathbf{x}|\mathbf{y})] + \mathbb{E}_{\mathbf{x}\sim q}[\log(1 - D(\mathbf{x}|\mathbf{y}))].
    \end{split}
\end{equation}

\noindent or, equivalently,
\begin{equation}\label{eq:cgan-loss-z}
    \begin{split}
        \underset{D}{\min}\text{\hspace{0.5em}}&\underset{G}{\max}\text{\hspace{0.5em}}V(G,D) := \\ 
                                               &\mathbb{E}_{\mathbf{x}\sim p}[\log(D(\mathbf{x}|\mathbf{y})] + \mathbb{E}_{\mathbf{z}\sim p_{\mathbf{z}}}[\log(1 - D(G(\mathbf{z}|\mathbf{y})))].
    \end{split}
\end{equation}

\subsection{Semi--Supervised Training with GANs}

GANs are most often used as generative models: after training, the discriminator is discarded, and the generator is used to generate synthetic samples that reflect the distribution of the source data; see Figure \ref{fig:generated-samples}. However, it is possible to modify the discriminator in the GAN by augmenting it witha network head that predicts the classification of the data, as illustrated in Figure \ref{fig:semi-supervised}. After training, the generator is discarded, and the discriminator can be used to classify samples from the source data. It has been shown that the semi--supervised training regime can be particularly effective in situations where the amount of training data is small \cite{salimans2016improved}.

\begin{figure}
    \vspace{-0.5\baselineskip}
    \centering
    \begin{subfigure}[t]{0.15\textwidth}
        \centering
        \includegraphics[width=0.98\textwidth]{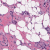}
    \end{subfigure}%
    \begin{subfigure}[t]{0.15\textwidth}
        \centering
        \includegraphics[width=0.98\textwidth]{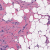}
    \end{subfigure}%
    \begin{subfigure}[t]{0.15\textwidth}
        \centering
        \includegraphics[width=0.98\textwidth]{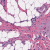}
    \end{subfigure}%

     \begin{subfigure}[t]{0.15\textwidth}
         \centering
         \includegraphics[width=0.98\textwidth]{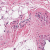}
     \end{subfigure}%
     \begin{subfigure}[t]{0.15\textwidth}
         \centering
         \includegraphics[width=0.98\textwidth]{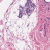}
     \end{subfigure}%
     \begin{subfigure}[t]{0.15\textwidth}
         \centering
         \includegraphics[width=0.98\textwidth]{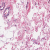}
     \end{subfigure}%

    \caption{Examples of the $50\times50$ crops used to train the model. The images in the top row are of IDC, while those in the second row are images of healthy tissue.}\label{fig:samples}
\end{figure}

\section{Methodology and Results}\label{section:method}

\subsection{Data} 

The data used for the experiments described here are the publicly available\footnote{\url{https://andrewjanowczyk.com/wp-static/IDC_regular_ps50_idx5.zip}} data first introduced in \cite{cruzroa2014automatic}.
These data consists of digitized histopathology slides from 162 women diagnosed with IDC at the Hospital of the University of Pennsylvania and The Cancer Institute of New Jersey. The slides were digitized via a whole--slide scanner at 40x magnification (0.25$\mu$m/pixel resolution), and each whole--slide image was downsampled by a factor of 16:1 to a resolution of 4$\mu$m/pixel. The ground truth annotations were obtained manually by an expert pathologist. The data were publicly released not in their original format, but rather as RGB patches of $50\times50$ pixels; see Figure \ref{fig:samples}. In total, the dataset contains 277,524 patches, of which 78,786, or 28\% are IDC, while the remaining 198,738 patches, or 72\% are healthy tissue. Note that the annotations were performed at 2x magnification or less, resulting in relatively coarse annotations, occasionally including some stromal or non--invasive tissue. 20\% of the data were held out for testing, and the model was trained on the remaining 80\% of the dataset.

    
\subsection{Architecture and Training}\label{subsection:architecture}

The model developed for these experiments is a conditional GAN based on the  DCGAN framework \cite{radford2015unsupervised}, where the generator is conditioned on the class of the input data, and the discriminator receives no conditioning, giving the modified value function
\begin{equation}\label{eq:cgan-loss-mine}
    \begin{split}
        \underset{D}{\min}\text{\hspace{0.5em}}&\underset{G}{\max}\text{\hspace{0.5em}}V(G,D) := \\
    &\mathbb{E}_{\mathbf{x}\sim p}[\log(D(\mathbf{x})] + \mathbb{E}_{\mathbf{x}\sim q}[\log(1 - D(\mathbf{x}))].
\end{split}
\end{equation}
\noindent or, equivalently,
\begin{equation}\label{eq:cgan-loss-z-mine}
    \begin{split}
        \underset{D}{\min}\text{\hspace{0.5em}}&\underset{G}{\max}\text{\hspace{0.5em}}V(G,D) := \\
                                               &\mathbb{E}_{\mathbf{x}\sim p}[\log(D(\mathbf{x}))] + \mathbb{E}_{\mathbf{z}\sim p_{\mathbf{z}}}[\log(1 - D(G(\mathbf{z}|\mathbf{y})))].
        \end{split}
\end{equation}
The generator uses a sequence of transposed convolutions to upsample the input latent vector, sometime referred to as a fully convolutional neural network \cite{long2015fully}. The discriminator is a convolutional neural network with two network heads, one that predicts the presence of IDC, and the other that predicts whether the observed data is real or synthetic. Both the generator and the discriminator use five transposed convolutional or convolutional layers with $3\times 3$ kernels. The number of filters in each convolutional layer of the discriminator was $64 \times layer \times \omega$, where $\omega$ is a width multiplier used to increase the capacity of the network; the number of filters in each transposed convolutional layer in the generator is calculated analogously, \emph{mutatis mutandis}. The generator uses ReLU activations, the discriminator uses leaky ReLU activations with $\epsilon=0.2$. No pooling layers where used in the discriminator; downsampling was accomplished by adjusting the stride of the convolutional layers as needed. The discriminator network heads consisted of a single fully connected layer. Spectral normalization was applied to all convolutional and transposed convolutional layers except the first and the last layers in the discriminator and gradient penalty was used to mitigate mode collapse \cite{miyato2018spectral,gulrajani2017improved}.

The network was trained for 200 epochs with minibatch size of 128 using the Adam optimizer ($\beta_1=0.5, \beta_2=0.999$) \cite{kingma2014adam}. The learning rate was fixed at $0.0002$ for the first 100 epochs, then reduced linearly to 0 for the remaining 100 epochs. The training data was augmented with vertically and horizontally flipped images. Traditional training loss curves tend to be uninformative when training GANs, so in addition to monitoring the generator and discriminator losses during training, samples from the generator outputs were periodically assessed qualitatively to insure that the generator was learning throughout the training process; samples generated by the generator are provided in Figure \ref{fig:generated-samples}. The model was implemented using the open--source machine learning framework PyTorch \cite{pytorch}. The model was trained on a workstation running Ubuntu 18.04 using two Titan Xp GPUs. Results are presented in Table \ref{table:results}.

\begin{table}[t]
    \centering
    \begin{tabular}{lcccc}
        \hline
        \multicolumn{1}{c}{\bf Metric} & \multicolumn{1}{c}{CNN (\cite{cruzroa2014automatic})} & \multicolumn{1}{c}{cGAN\_1} & \multicolumn{1}{c}{cGAN\_2} & \multicolumn{1}{c}{cGAN\_4} \\ \hline
        Accuracy & NA & 86.68\% & 87.45\% & \textbf{88.33}\% \\
        BAC & \textbf{84.23}\%& 81.15\% & 83.19\% & 83.54\% \\
		Precision & 65.40\% & 81.94\% & 80.85\% & \textbf{84.39}\% \\ 
        Recall & \textbf{79.60}\% & 68.29\% & 73.29\% & 72.41\% \\
        Specificity & NA & 94.00\% & 93.09\% & \textbf{94.66}\% \\
        F1 & 71.80\% & 74.50\% & 76.88\% & \textbf{77.94}\% \\
        \hline
    \end{tabular}
    \caption{Results from semi--supervised experiments. In expressions of the form cGAN\_$\omega$, the value $\omega$ is the width multiplier described in Section \ref{subsection:architecture}.}\label{table:results}
\end{table}

\section{Conclusions and Future Work}\label{section:conclusions}

In this paper we present the results of an investigation into the use of GANs and conditional GANs for automatic detection of IDC in breast histopathology images. The advantages of a GAN or cGAN framework is that the generator in the framework learns during the training process to generate data that follows the distribution of the training data, thus supplementing the training dataset with additional high--quality synthetic training data. These models achieve high accuracy, precision, specficity, and F1--scores, and competitive balanced accuracy scores, while being less sensitive than a conventional convolutional neural network model.

There are several avenues for future work in this vein. One of the advantages of semi--supervised GAN training is that in situations with limited data, it is often possible to achieve superior performance over other methods on similarly sized data. As noted in \cite{mahmood2019structured}, most GAN discriminators are rather shallow in comparison to modern classifier architectures. Semi--supervised training with a fixed dataset may allow one to increase the capacity of the discriminator over a base classifier CNN and thereby improve performance beyond the results described here.

The ability to condition the generator of a conditional GAN on some supplementary data, such as the class of the data to be generated, is a noteworthy aspect of this model. Future investigations will explore other conditioning approaches; one possibility, for example, is to condition the generator based on both the class to be generated as well as the location of the patch in the whole slide image.

The algorithm described here has relatively high precision, but lower recall/sensitivity than other automatic detection methods based on CNNs. Increasing the recall of the algorithm while maintaining the precision, perhaps by weighting the loss function, is another potentially fruitful avenue for future work. Finally, in a purely theoretical direction, there is still much work to be done to understand the complex interplay between adverarial and classification loss in semi--supervised GAN training.

\bibliographystyle{IEEEbib}
\bibliography{breast-histopathology}

\end{document}